\documentclass[aps,prb,floatfix,amsmath,preprint,eqsecnum,nofootinbib]{revtex4}
\usepackage{graphicx}
\usepackage{dcolumn}
\usepackage{bm}
\usepackage{setspace}   

\usepackage{amsmath,amssymb,bm}
\usepackage{latexsym}
\usepackage{ifpdf}
\ifpdf
\fi
\begin{document}
\title{Magnetization of the Shastry-Sutherland antiferromagnet\\ 
near the Ising limit}
\author{Fangzhou Liu}
\affiliation{Department of Physics, Harvard University, Cambridge MA 02138, USA}
\author{Subir Sachdev}
\affiliation{Department of Physics, Harvard University, Cambridge MA 02138, USA}
\date{\today\\
\vspace{1.6in}}
\begin{abstract}
Motivated by recent experiments on TmB$_4$ (Phys. Rev. Lett. {\bf 101}, 177201 (2008)),
we examine the phase diagram of the Shastry-Sutherland antiferromagnet in an applied
magnetic field in the limit of strong Ising anisotropy. In classical Ising limit, we demonstrate
that the only fractional magnetization plateau is at 1/3 of the saturated magnetization.
We study the perturbative influence of transverse quantum spin fluctuations, and present evidence that
they can stabilize a narrow 1/2 magnetization plateau.
\end{abstract}

\maketitle

\section{Introduction}

The Shastry-Sutherland antiferromagnet \cite{ssafm} is a rare example of an exactly soluble frustrated spin
model in two spatial dimensions. Its ground state is a direct product of near-neighbor spin-singlet pairs, and there is an energy gap to all excitations. Much theoretical and attention has been lavished on this model after the discovery \cite{scbo1,scbo2}
of its experimental realization in $\mathrm{SrCu_2(BO_3)_2}$.

This paper will examine the phases of the Shastry-Sutherland antiferromagnet for the case
of exchange interactions with a strong Ising anisotropy, and in the presence of an applied magnetic field.
This is motivated by a recent experimental study \cite{experimental} of $\mathrm{TmB_4}$, in which the magnetic
Tm ions are believed to be Ising-like.

The experimental studies on the compound $\mathrm{TmB_4}$ determined both on its phase diagram as a function of temperature,
and the magnetization diagram\cite{experimental} as a function of the external magnet field. In the magnetization diagram, as the external magnetic field increased from 0, magnetization plateaus with $M/M_{sat}=1/7,1/8,1/9...$ are seen, followed by a major 1/2 magnetization plateau before reaching saturation\cite{experimental}.

In this paper a classical Ising model was first used in Monte Carlo calculation to simulate the magnetization process. Based on the simulation result, a more detailed analysis was done on the ground state phase diagram for Ising-like Shastry-Sutherland (SS) lattice. 
Going beyond the classical Ising limit, we accounted for the quantum effects of the transverse spin interaction by the effective Hamiltonian method of Sen, Wang and Damle\cite{Isingsubspace}; this is valid in the limit of large spin $S$. We will show
that this effective Hamiltonian does contain terms which stabilize a 1/2 magnetization plateau.

\begin{figure}
\includegraphics[width=1.5in]{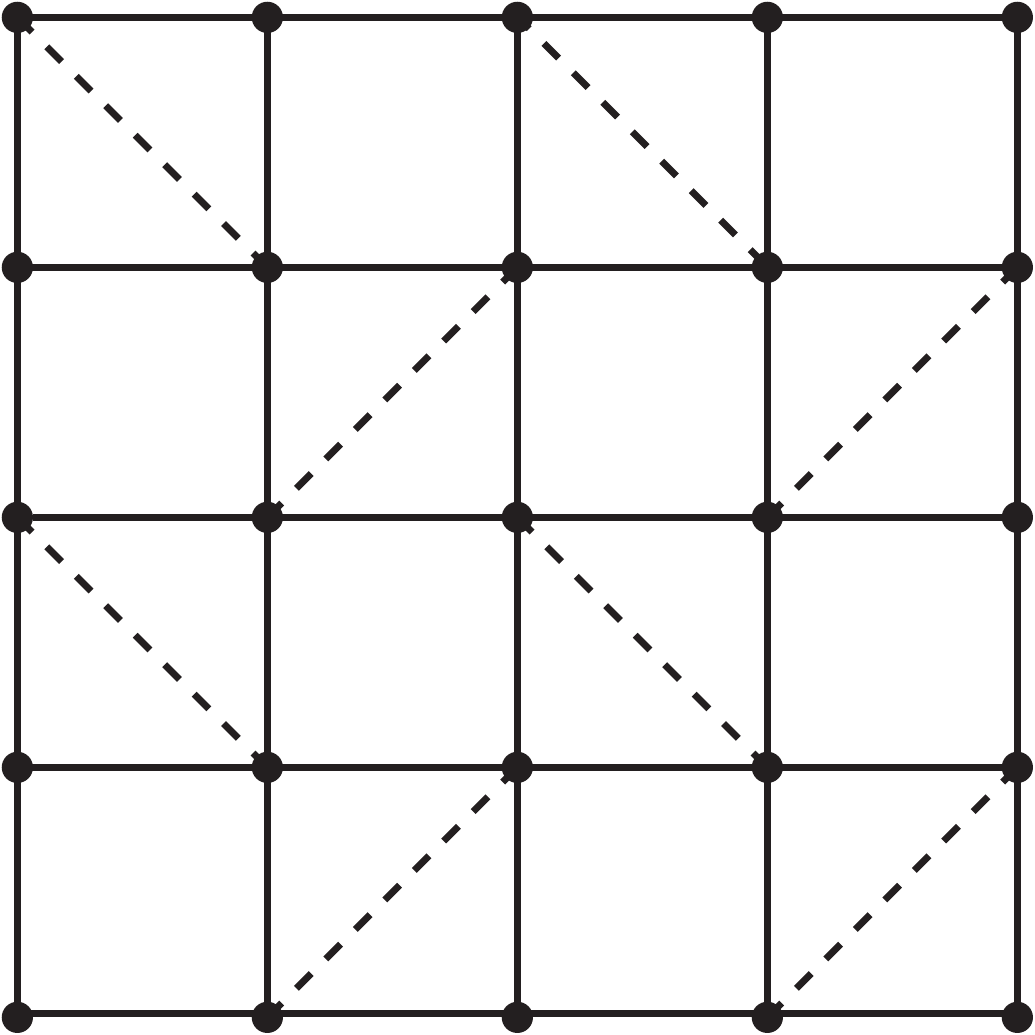}
\caption{The Shastry-Sutherland lattice. The exchange $J_1$ acts
between sites separated by the horizontal and vertical links,
which the exchange $J_2$ acts across the diagonal links.\label{fig:ssl}}
\end{figure}

The theoretical models in this paper were all based on the two-dimensional lattice, with magnetic ion located on each site of the SS lattice. The magnetic ion interacts only with the nearest neighbors through exchange interactions, with square bond strength $J_{1}$ and diagonal bond $J_{2}$ (see Fig.~\ref{fig:ssl}).

While this work was being completed, we learnt of the work of Meng and Wessel \cite{meng}; where their results
overlap with ours, they are in agreement.

\section{The SSL Ground State Phase Diagram in the Ising Model}

A naive approach to the problem is the classic spin-$1/2$ Ising model. Spins, which can take two different values (up or down), are located at each corner of the SS lattice, with anti-ferromagnetic exchange interaction $J_{1}>0$ and $J_{2}>0$ between the nearest neighbors. To simulate the magnetization process, the Monte Carlo Method was used in the simulation. The simple Hamiltonian was taken as:
\begin{equation}
\label{Ising Hamiltonian}
H=J_1\sum_{J_1\langle i,j \rangle}\sigma_i\sigma_j+J_2\sum_{J_2\langle i,j \rangle}\sigma_i\sigma_j-\mu B\sum_i\sigma_i
\end{equation}
where the first summation runs over all the square bonds and the second runs over all the diagonal bonds, $\sigma$ denotes the spins with $\sigma_i=\pm1$, and $\mu B$ term denotes the energy gained from the external magnetic field.

The detailed Monte Carlo process was done in the following way. The lattice size was chosen to be $18\times18$, which is a common multiple of both 2 and 3 (the reasons will follow), with periodic boundary condition at the four sides. Periodic boundary condition in SS lattice requires the length to be a multiple of 2. In the Metropolis algorithm, iteration times before reaching equilibrium and after reaching equilibrium were all set to be 100,000 times, the temperature was set very close to zero, and the bond strength $J_1$ was set to one.

\begin{figure}
\includegraphics{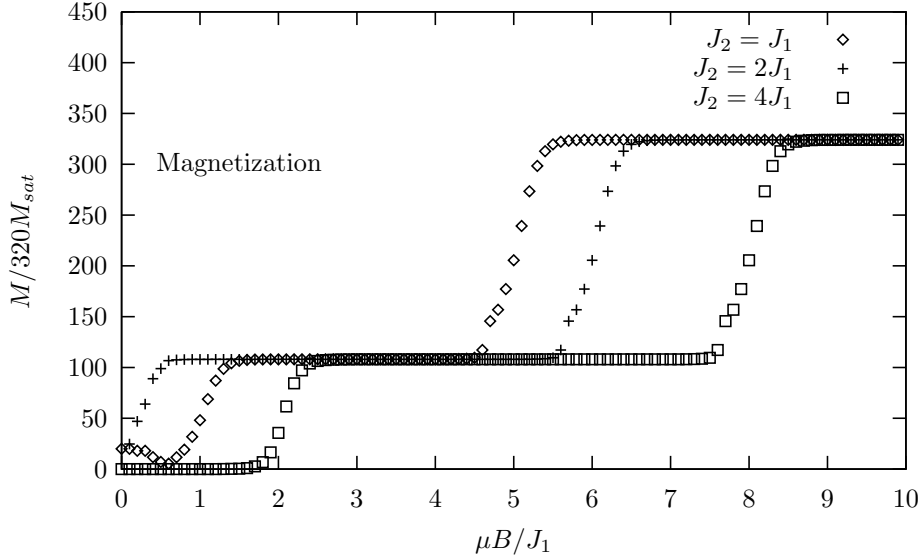}
\caption{The magnetization process when $J_2$ take the three different values: $J_2=J_1, J_2=2J_1, J_2=4J_1$. \label{magnetization}}
\end{figure}
\begin{figure}
\includegraphics{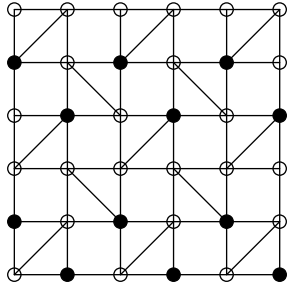}
\caption{The spin configuration pattern obtained in the one-third magnetization plateau.}
\label{1/3 plateau spin config.}
\end{figure}

The results of the simulation are shown in Fig.~\ref{magnetization}, where different lines represent different values of $J_2/J_1$ ratio. As can be seen from the figure, for all values of the $J_2/J_1$ ratio, there's always a very large $1/3$ magnetization plateau, and no $1/2$ plateau is observed. It will be shown later in this section that no $1/2$ plateau is possible at zero temperature in the Ising model. Similar simulation results were also be obtained when lattice size was chosen to be 12 or 24 (again, both as integer multiple of 2 and 3). It is important to note that in this simulation, the choice of the lattice size matters, and a wrong choice of lattice size could lead to different conclusions. To understand the choice of lattice size, the resulting spin configuration in our $1/3$ magnetization plateau is shown here in Fig.~\ref{1/3 plateau spin config.}.

In Fig.~\ref{1/3 plateau spin config.}, empty dots represent spin ups while black dots represent spin downs. In the figure, $1/3$ of the total spins are pointing down while others pointing up. Also, every spin down is surrounded by spin ups both through $J_1$ bond and $J_2$ bond. This is the key feature of this spin configuration, which makes its energy the lowest among all that have $1/3$ of spins pointing down (c.f. Eq.~(\ref{Ising Hamiltonian})). Due to the period of 3 in this spin configuration (a repetition of three rows in the figure), the lattice size has to be a multiple of 3 to make this configuration possible. Note that the degeneracy of this spin configuration has a very limited value of 6, compared to the large degeneracy of other types. From the figure, we can now write down the Hamiltonian for this particular spin configuration (Hamiltonian for the $1/3$-plateau phase) by using Eq.~(\ref{Ising Hamiltonian}):
\begin{equation}
\label{1/3 plateau Hamiltonian}
H=-\frac{2}{3}NJ_1-\frac{1}{6}NJ_2-\frac{1}{3}N\mu B
\end{equation}
where $N$ represents the total number of lattice sites, and other parameters are the same as Eq.~(\ref{Ising Hamiltonian}).

\begin{figure}
\includegraphics{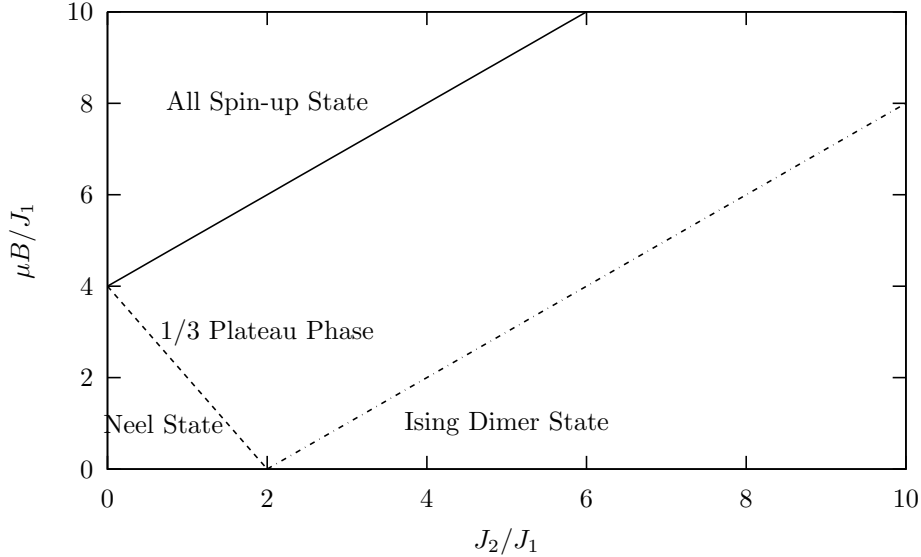}
\caption{The ground state phase diagram for SS lattice, with $J_2/J_1$ and $\mu B/J_1$ being two independent variables. The boundaries are obtained simply by comparing the Hamiltonian of different phases using Eq.~(\ref{Ising Hamiltonian}).}
\label{Ising phase diagram}
\end{figure}

We are now at the position of giving out a full ground state phase diagram. The proposed ground state phase diagram is shown here in Fig.~\ref{Ising phase diagram}.
Different phases shown are the different zero-temperature ground states of the general SS lattice with Hamiltonian of Eq.~(\ref{Ising Hamiltonian}). As shown in the figure, when the external field is zero, we have the Neel State (opposite spins on square bonds) when $J_2/J_1\ll1$, and the Ising Dimer State (opposite spins on diagonal bonds) when $J_2/J_1\gg1$. As the external field increases, the $1/3$ plateau phase will arise, giving us a positive total magnetization. When the external field is large enough, the All Spin-up State will show up, giving us the magnetization saturation.

Further reflection explains the existence of $1/3$ plateau: if we rewrite the Hamiltonian as a function of the ratio between spin-down-number and total-spin-number (denoted by $\rho$), the $1/3$ plateau phase will have the largest $\rho$ value among all states that have the key feature (all the spin downs are surrounded by spin ups through both $J_1$ and $J_2$ bonds). That is to say, the $1/3$ plateau state will have the largest number of spin downs while keeping the key feature. This will make the 1/3 plateau phase an extreme. Different extremes of the Hamiltonian are obtained at $\rho=0$(all spin-up state), $\rho=1/3$(1/3 plateau phase) and $\rho=1/2$ (Neel state and Ising dimer state), which are exactly the different phases shown in Fig.~\ref{Ising phase diagram}.

To demonstrate the reliability of the proposed ground state phase diagram, an exhaustive computational simulation was done on a $4\times6$ lattice, by going through all the possible spin configurations and picking out the ones with the lowest energy (altogether $2^{24}\approx1.68\times10^{7}$ configurations). The result of the simulation agreed completely with Fig.~\ref{Ising phase diagram}.

It's also very instructive to look back and compare Fig.~\ref{Ising phase diagram} with the magnetization result obtained in Fig.~\ref{magnetization}. The two figures agree well at the critical points when phase transition occurs from the 1/3 plateau to the all spin-up state, with $J_2/J_1$ value being 1, 2, 4, and $\mu B/J_1$ value being 5, 6, 8, respectively.

To conclude, as a direct application of the obtained phase diagram, we'll present here a short proof showing that no $1/2$ plateau ground state is possible at zero temperature in the Ising model. This is achieved by showing that there's always a lower energy state than any possible states with a 1/2-saturation magnetization. First, in view of Eq.~(\ref{Ising Hamiltonian}), it's not hard to get the lowest energy expression for states with a 1/2 magnetization:
\begin{equation}
\label{1/2 plateau Hamiltonian}
H=-\frac{1}{2}N\mu B
\end{equation}
In the above equation, by minimizing the energy expression, both $J_1$ and $J_2$ terms accidentally vanished, leaving only the $\mu B$ term. The minimizing process could be easily understood as follows: for states with a 1/2-saturation magnetization, the total number of spin downs would be $1/4N$($N$ being the total site number), so that the maximum number of antiparallel square-bond pairs would be $1/4N \times 4 = N$. Noting that there're altogether $2N$ square-bonds, the antiparallel pairs and parallel pairs would cancel each other in the $J_1$ term. Similar analysis applies to the $J_2$ term.

The energy expression for the All Spin-up State is also needed for the proof:
\begin{equation}
\label{all spin-up Hamiltonian}
H=2NJ_1+\frac{N}{2}J_2-N\mu B
\end{equation}
Now, by direct comparison of the three energy expressions (Eqs. \ref{1/3 plateau Hamiltonian}, \ref{1/2 plateau Hamiltonian} and \ref{all spin-up Hamiltonian}), we see that either the All Spin-up State (Eq.~(\ref{all spin-up Hamiltonian})) or the 1/3 Plateau Phase (Eq.~(\ref{1/3 plateau Hamiltonian})) have lower energy than the minimum energy we obtained for states with 1/2-saturation magnetization (Eq.~(\ref{1/2 plateau Hamiltonian})) throughout the phase diagram. That is to say, states with 1/2-saturation magnetization could never have the lowest energy. Thus no 1/2 magnetization plateau is possible in the Ising limit at zero temperature.

\section{Perturbative Influence: Transverse quantum fluctuation}

As the classical Ising model does not give us an understanding of the major 1/2 plateau observed in the experiment, \cite{experimental} improvements were made to our Ising model Hamiltonian. 
In this section, we focus on the perturbative effects of the transverse fluctuation on the SSL spin multiplets with large easy axis anisotropy:
\begin{align}
\nonumber
H=&-D\sum_i(S_{i}^z)^2+J_1\sum_{J_1\langle i,j \rangle}S_i^zS_j^z+J_2\sum_{J_2\langle i,j \rangle}S_i^zS_j^z-\mu B\sum_iS_i^z
\\\label{easy axis Hamiltonian}
&+p \left[ J_1\sum_{J_1\langle i,j \rangle}(S_i^xS_j^x+S_i^yS_j^y)+J_2\sum_{J_2\langle i,j \rangle}(S_i^xS_j^x+S_i^yS_j^y) \right]
\end{align}
In the above equation, $D$ is the strength of the easy axis anisotropy along the $z$ axis, $S_{i}$ are the spin multiplets, $J_1$ and $J_2$ are the bond strengths same as in Eq.~(\ref{Ising Hamiltonian}), and the $p$ term was introduced to take care of the transverse interaction, with parameter $p$ adjustable to count for the asymmetry between $z$ axis and the $x,y$ axis.

As argued in Ref.~\onlinecite{Isingsubspace}, since the problem turns out to be a strong Ising anisotropy problem (see Ref.~\onlinecite{experimental}), a $J/D$ expansion of the Hamiltonian will give us the leading quantum effects. For the perturbation process, following the method introduced by Ref.~\onlinecite{Isingsubspace}, we first split the Hamiltonian as $H=H_0+H'$, in which
\begin{equation}
\label{H-nord}
H_0=-D\sum_i(S_{i}^z)^2+J_1\sum_{J_1\langle i,j \rangle}S_i^zS_j^z+J_2\sum_{J_2\langle i,j \rangle}S_i^zS_j^z-\mu B\sum_iS_i^z
\end{equation}
\begin{equation}
\label{H-prime}
H'=\frac{1}{2}p(J_1\sum_{J_1\langle i,j \rangle}(S_i^+S_j^-+S_i^-S_j^+)+J_2\sum_{J_2\langle i,j \rangle}(S_i^+S_j^-+S_i^-S_j^+))
\end{equation}
Note that $H'$ is actually the $p$ term in Eq.~(\ref{easy axis Hamiltonian}). When $D$ is large, the ground state manifold of $H_0$ is just the Ising limit we adopted in the previous section, where the spins are all maximally polarized along the $z$ axis (so that $S_i^z=S\sigma_i$). This ground state manifold is referred to as the Ising subspace. Degenerate perturbation theory will now be applied by treating $H'$ as a perturbation on this $H_0$ ground state manifold (or, the Ising subspace).

It's not hard to check that both the first-order perturbation terms and the second-order off-diagonal perturbation terms vanish, leaving only the second-order diagonal terms:
\begin{equation}
\label{H-(2)}
H^{(2)}=\sum_{k\notin G}\frac{|
\langle 0|H'|k \rangle |^2}{E_0-E_k}
\end{equation}
where $G$ denotes the Ising subspace, $0$ denotes any state in the Ising subspace, $k$ denotes any state outside the Ising subspace, and $E_0$ and $E_k$ are the energies for the corresponding states. For further convenience, rewrite the above expression in the following way:
\begin{equation}
\label{Two-term H-(2)}
H^{(2)}=\sum_{k_1\notin G}\frac{|\langle0|H'|k_1\rangle |^2}{E_0-E_{k_1}}
+\sum_{k_2\notin G}\frac{|\langle0|H'|k_2\rangle |^2}{E_0-E_{k_2}}
\end{equation}
where $k_1$ are the states that, compared with states in the Ising subspace, have different spins only at the two ends of one single square bond, while $k_2$ having different spins only at the two ends of one single diagonal bond (it is not hard to see that $k_1$ and $k_2$ will include all the states with non-vanishing value in Eq.~(\ref{H-(2)})).

After a somewhat lengthy calculation of the above equation by calculating the two terms separately (note that in the calculation, the two terms would have different energy denominators), and by performing a small-$J/D$ expansion for both terms, we 
finally arrive at the resulting perturbation term:
\begin{align}
\nonumber
H^{(2)}=&-\frac{p^{2}J_{1}^{2}S^{2}}{2D(2S-1)}\sum_{J_{1}\langle i,j \rangle}\frac{1-\sigma_{i}\sigma_{j}}{2}\left[1+\frac{J_{1}}{2D(2S-1)}+
\frac{J_{1}S(\sigma_{i}H_{i}+\sigma_{j}H_{j})}{2D(2S-1)}\right.
\\\nonumber
&+\left.\frac{J_{2}S(\sigma_{i}\sigma_{id}+\sigma_{j}\sigma_{jd})}{2D(2S-1)}\right]
\\\label{perturbation Hamiltonian}
&-\frac{p^{2}J_{2}^{2}S^{2}}{2D(2S-1)}\sum_{J_{2}\langle i,j \rangle}\frac{1-\sigma_{i}\sigma_{j}}{2}\left[1-\frac{J_{2}}{2D}+
\frac{J_{1}S(\sigma_{i}H_i+\sigma_jH_j)}{2D(2S-1)}\right]
\end{align}
and the effective Hamiltonian in the Ising subspace was obtained as:
\begin{equation}
\label{total Hamiltonian}
\tilde{H}=H_{0}+H^{(2)}
\end{equation}
In Eq.~(\ref{perturbation Hamiltonian}), $S$ is the absolute value of the spin multiplet, $\sigma_i=\pm1$ are the spin-ups or spin-downs as in the previous Ising limit approach, whereas $H_i=\sum_i\alpha_{ij}\sigma_j$ is the local field, with $\alpha_{ij}=1$ for $\sigma_i$'s square-bond neighbors and $\alpha_{ij}=0$ otherwise; and $\sigma_{id}=\alpha_{ij}\sigma_j$, with $\alpha_{ij}=1$ for $\sigma_i$'s diagonal bond neighbor and $\alpha_{ij}=0$ otherwise.

\subsection{The New Phase Diagram}
With the new effective Hamiltonian obtained above, we can now propose a new ground state phase diagram for the system. Note that this is far from a rigorous proof of the diagram, as the Hamiltonian is much more complex than in the Ising model, and there's now no direct way to get the global energy minimum. Also, the exhaustive simulation is impossible in this case. Yet as the new Hamiltonian is based on the original Ising Hamiltonian, it's reasonable to start from the ground states in the original phase diagram.

\begin{figure}
\includegraphics{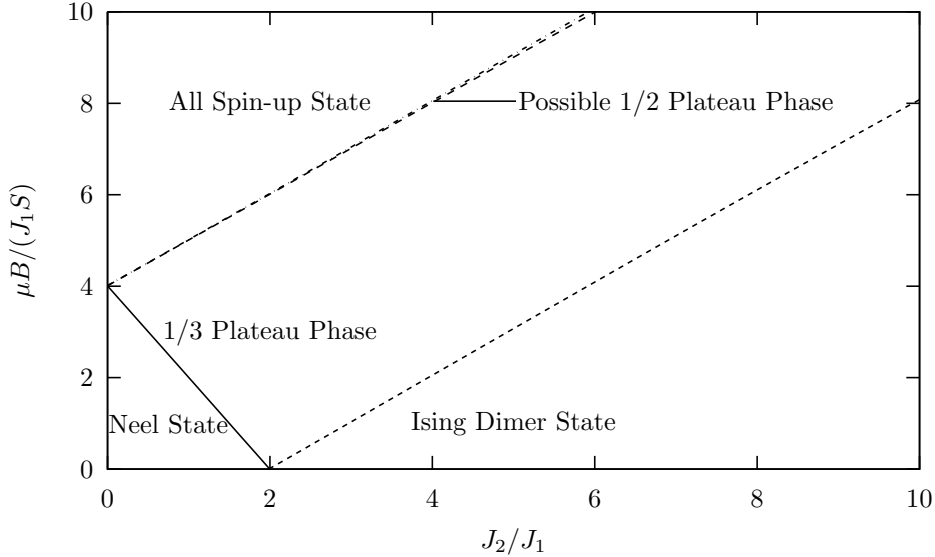}
\caption{The possible ``new" phase diagram with the new effective Hamiltonian. In the diagram, all the boundaries between different phases are obtained using the effective Hamiltonian. Note that in the figure, there's now a split between the ``All Spin-up State" and the ``1/3 Plateau Phase", where a state with magnetization being 1/2 of the saturation value could be proved to have lower energy than both of the two states, and thus provide a region for a possible ``1/2 Plateau Phase".\label{newphasediagram1}}
\end{figure}
In Fig.~\ref{newphasediagram1}, we directly adopted the previous four ground states without rigorous proof. The boundaries in the figure are all obtained using the new Hamiltonian, where we've set the magnitude of the spin multiplet $S$ to be $6$ (as in the experiment\cite{experimental}), the $J_1/D$ value to be $1/5$, and the parameter $p$ to be $1$. For the convenience of direct comparison with the original phase diagram, the unit of the vertical axis was divided by a factor of $S$. Comparison with the original phase diagram (Fig.~\ref{Ising phase diagram}) shows their similarities, which should be the case.

\begin{figure}
\includegraphics{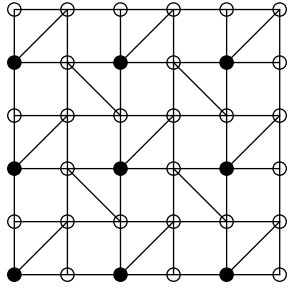}
\caption{The spin configuration pattern of the ``new possible ground state" found. As could be seen in the figure, 1/4 of the total spins are pointing down, making the total magnetization 1/2 of the saturation value.\label{1/2 plateau spin config.}}
\end{figure}

\begin{figure}
\includegraphics{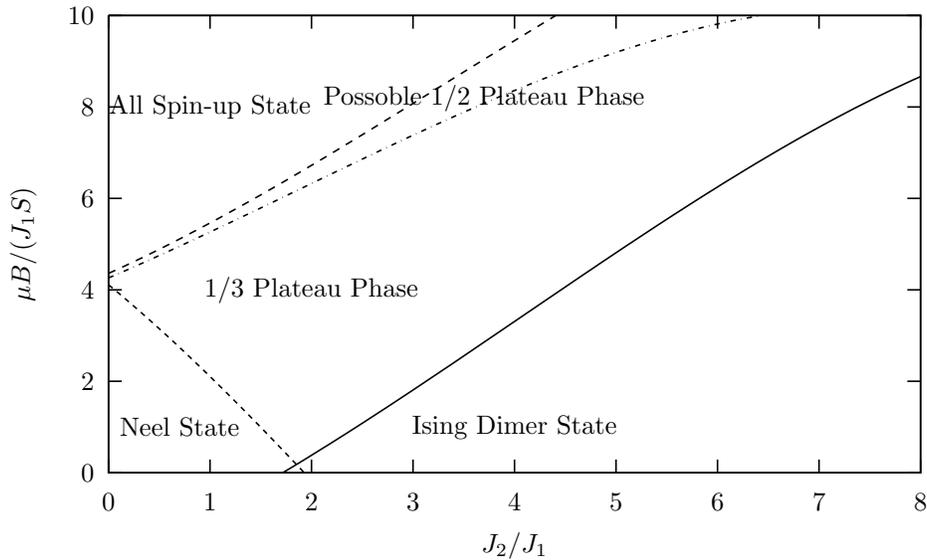}
\caption{The phase diagram obtained with a more dramatic $p$ value. The parameter $p$ is changed to 5 in the figure, to make the possible ``1/2 Plateau Phase" more obvious to see. Note that in the figure, the boundaries began to cross each other, which means that the boundary behavior of the phase diagram is more complex than before, and further analysis are needed.\label{newphasediagram2}}
\end{figure}

Of course, minor corrections are necessary for the new phase diagram, especially at the boundaries between the four ``old" ground states; here presented in Fig.~\ref{newphasediagram1} is only a correction we found at the boundary between ``All Spin-up State" and the ``1/3 Plateau Phase". As could be seen in the figure, the boundary now split up, and a new state with magnetization being 1/2-saturation is found to have lower energy. Its detailed spin configuration pattern is showed here in Fig.~\ref{1/2 plateau spin config.} (again, black dots represent spin downs while empty dots ups).
Although it's impossible to prove that this configuration has the lowest energy, this interesting finding at least gives a region where the ``1/2 magnetization plateau" is possible. If we redraw a diagram of magnetization against the external field, before the magnetization reach its saturation value, a 1/2 plateau may occur. To make this region more dramatic and easy to see in the phase diagram, the parameter $p$ was tuned to be $5$ in Fig.~\ref{newphasediagram2}.

At the end, it's also reasonable to further suspect that the small fractional magnetization plateaus with $M/M_{sat}=1/7,1/8,1/9...$ found in the experiment \cite{experimental} could also be possibly explained by similar arguments. As the boundary between ``Neel State" and ``1/3 Plateau Phase" start to split up, fractional plateau phase may have lower energy and arise from the effective Hamiltonian correction.

\section{Conclusion}
In this paper, a thorough analysis was done for the two dimensional SS lattice with the Ising model, giving us the full ground state phase diagram (see Fig.~\ref{Ising phase diagram}). The phase diagram is consistent with both the Monte Carlo simulations and the exhaustive simulation for a small system. As shown in the diagram, except for the Neel State, Ising Dimer State and the All Spin-up State, only a 1/3 magnetization plateau phase could be found, and no 1/2 magnetization plateau is possible in the Ising limit. Further study was done by considering the perturbative effects of the transverse fluctuations, followed by $J/D$ strong anisotropy expansion. The new effective Hamiltonian showed minor corrections to the phase diagram, and more importantly, it gave a possible region in the phase diagram where 1/2 magnetization plateau may occur. Further analysis could be done in the future to confirm this ``1/2 plateau region", or to probably explore corrections at other boundaries, in the hope of finding other small ``fractional plateau phases" in this model. We hope that the work done would be useful for further understanding of the behaviors of $\mathrm TmB_4$, and for the further study of all the SSL quantum magnets.

\acknowledgements

This research was
supported by the NSF under grant DMR-0757145.

\end{document}